\magnification 1200

%
% ??/epsf.tex (written by Radical Eye Software and copied below)
% defines the macro \epsfbox with one argument,
% the encapsulated PostScript file to include.
% Invoking it causes a \vbox with the natural size of the drawing
% to be inserted at the point of invocation.
% Usually figures are meant to be centered and set off, and possibly
% to have a title and/or a figure number. The macros below do that.
% You should assign values that please you to the variables
% \abovefigskip, \belowfigskip, figtitleskip, figtitlefont,...
% AFTER \inputting the present file: \input /u2/kbi/tex/epsfig
% and BEFORE the first invocation of any of the macros below.
% RESET AT YOUR PLEASURE THE VARIABLES AT THE VERY BOTTOM!
%
% For convenience we make a dimension for figures:
\newdimen\FigSize       \FigSize=.9\hsize % alter at your convenience
%
% For a SCALED HORIZONTALLY CENTERED FIGURE use \epsfig.
% First argument is the horizontal width of the figure, given
% in any way TeX can understand.
% The second argument is an encapsulated PostScript file name(filnam.eps).
% Note the mandatory semicolons between arguments in this example!:
% \epsfig .8\hsize; example.ps;
% will put a centered scaled \vbox of width .8\hsize suitably offset
% at the point of invocation
\newskip\abovefigskip   \newskip\belowfigskip
\gdef\epsfig#1;#2;{\par\vskip\abovefigskip\penalty -500
   {\everypar={}\epsfxsize=#1\nd \centerline{\epsfbox{#2}}}%
    \vskip\belowfigskip}%
%
% SCALED TITLED EPSFIG HORIZONTALLY CENTERED: \tepsfig.
% First argument is the horizontal width of the figure,
% second an encapsulated PostScript file name,
% third a title for the figure.
% Note the mandatory semicolons between arguments!
% example: \tepsfig5truein; example.ps;{This is a figure}
\newskip\figtitleskip
\gdef\tepsfig#1;#2;#3{\par\vskip\abovefigskip\penalty -500
   {\everypar={}\epsfxsize=#1\nd
    \vbox
      {\centerline{\epsfbox{#2}}\vskip\figtitleskip
       \centerline{\figtitlefont#3}}}%
    \vskip\belowfigskip}%
%
% SCALED NUMBERED TITLED EPSFIG HORIZONTALLY CENTERED: \nepsfig
% The figure number is automatically increased for every
% invocation of \nepsfig or \nipsfig.
\newcount\FigNr \global\FigNr=0
\gdef\nepsfig#1;#2;#3{\global\advance\FigNr by 1
   \tepsfig#1;#2;{Figure\space\the\FigNr.\space#3}}%
%
%
% Often you would rather have TeX decide where to put the figure
% by using \midinsert. Here are macros that do that
% (mnemonics ``ipsfig'' is for ``midInsert PS FIGure'')
%
% TeX-PLACED SCALED EPSFIG HORIZONTALLY CENTERED: \ipsfig
\gdef\ipsfig#1;#2;{%\goodbreak ??
   \midinsert{\everypar={}\epsfxsize=#1\nd
              \centerline{\epsfbox{#2}}}
   \endinsert}%
%
% TeX-PLACED SCALED TITLED EPSFIG HORIZONTALLY CENTERED: \tipsfig
\gdef\tipsfig#1;#2;#3{\midinsert
   {\everypar={}\epsfxsize=#1\nd
    \vbox{\centerline{\epsfbox{#2}}%
          \vskip\figtitleskip
          \centerline{\figtitlefont#3}}}\endinsert}%
%
% TeX-PLACED SCALED NUMBERED TITLED EPSFIG HORIZONTALLY CENTERED: \nipsfig
% example: \nipsfigd.9\hsize;example.ps;{This is an example figure}
\gdef\nipsfig#1;#2;#3{\global\advance\FigNr by1%
  \tipsfig#1;#2;{Figure\space\the\FigNr.\space#3}}%
\newread\epsffilein    % file to \read
\newif\ifepsffileok    % continue looking for the bounding box?
\newif\ifepsfbbfound   % success?
\newif\ifepsfverbose   % report what you're making?
\newdimen\epsfxsize    % horizontal size after scaling
\newdimen\epsfysize    % vertical size after scaling
\newdimen\epsftsize    % horizontal size before scaling
\newdimen\epsfrsize    % vertical size before scaling
\newdimen\epsftmp      % register for arithmetic manipulation
\newdimen\pspoints     % conversion factor
\pspoints=1bp          % Adobe points are `big'
\epsfxsize=0pt         % Default value, means `use natural size'
\epsfysize=0pt         % ditto
\def\epsfbox#1{\global\def\epsfllx{72}\global\def\epsflly{72}%
   \global\def\epsfurx{540}\global\def\epsfury{720}%
   \def\lbracket{[}\def\testit{#1}\ifx\testit\lbracket
   \let\next=\epsfgetlitbb\else\let\next=\epsfnormal\fi\next{#1}}%
\def\epsfgetlitbb#1#2 #3 #4 #5]#6{\epsfgrab #2 #3 #4 #5 .\\%
   \epsfsetgraph{#6}}%
\def\epsfnormal#1{\epsfgetbb{#1}\epsfsetgraph{#1}}%
\def\epsfgetbb#1{%
%
%   The first thing we need to do is to open the
%   PostScript file, if possible.
%
\openin\epsffilein=#1
\ifeof\epsffilein\errmessage{I couldn't open #1, will ignore it}\else
%
%   Okay, we got it. Now we'll scan lines until we find one that doesn't
%   start with %. We're looking for the bounding box comment.
%
   {\epsffileoktrue \chardef\other=12
    \def\do##1{\catcode`##1=\other}\dospecials \catcode`\ =10
    \loop
       \read\epsffilein to \epsffileline
       \ifeof\epsffilein\epsffileokfalse\else
%
%   We check to see if the first character is a % sign;
%   if not, we stop reading (unless the line was entirely blank);
%   if so, we look further and stop only if the line begins with
%   `%%BoundingBox:'.
%
          \expandafter\epsfaux\epsffileline:. \\%
       \fi
   \ifepsffileok\repeat
   \ifepsfbbfound\else
    \ifepsfverbose\message{No bounding box comment in #1; using
defaults}\fi\fi
   }\closein\epsffilein\fi}%
%
%   Now we have to calculate the scale and offset values to use.
%   First we compute the natural sizes.
%
\def\epsfsetgraph#1{%
   \epsfrsize=\epsfury\pspoints
   \advance\epsfrsize by-\epsflly\pspoints
   \epsftsize=\epsfurx\pspoints
   \advance\epsftsize by-\epsfllx\pspoints
%
%   If `epsfxsize' is 0, we default to the natural size of the picture.
%   Otherwise we scale the graph to be \epsfxsize wide.
%
   \epsfxsize\epsfsize\epsftsize\epsfrsize
   \ifnum\epsfxsize=0 \ifnum\epsfysize=0
      \epsfxsize=\epsftsize \epsfysize=\epsfrsize
%
%   We have a sticky problem here:  TeX doesn't do floating point
arithmetic!
%   Our goal is to compute y = rx/t. The following loop does this reasonably
%   fast, with an error of at most about 16 sp (about 1/4000 pt).
%
     \else\epsftmp=\epsftsize \divide\epsftmp\epsfrsize
       \epsfxsize=\epsfysize \multiply\epsfxsize\epsftmp
       \multiply\epsftmp\epsfrsize \advance\epsftsize-\epsftmp
       \epsftmp=\epsfysize
       \loop \advance\epsftsize\epsftsize \divide\epsftmp 2
       \ifnum\epsftmp>0
          \ifnum\epsftsize<\epsfrsize\else
             \advance\epsftsize-\epsfrsize \advance\epsfxsize\epsftmp
\fi
       \repeat
     \fi
   \else\epsftmp=\epsfrsize \divide\epsftmp\epsftsize
     \epsfysize=\epsfxsize \multiply\epsfysize\epsftmp
     \multiply\epsftmp\epsftsize \advance\epsfrsize-\epsftmp
     \epsftmp=\epsfxsize
     \loop \advance\epsfrsize\epsfrsize \divide\epsftmp 2
     \ifnum\epsftmp>0
        \ifnum\epsfrsize<\epsftsize\else
           \advance\epsfrsize-\epsftsize \advance\epsfysize\epsftmp \fi
     \repeat
   \fi
%
%  Finally, we make the vbox and stick in a \special that dvips can parse.
%
   \ifepsfverbose\message{#1: width=\the\epsfxsize,
height=\the\epsfysize}\fi
   \epsftmp=10\epsfxsize \divide\epsftmp\pspoints
   \vbox to\epsfysize{\vfil\hbox to\epsfxsize{%
      \includegraphics{#1}%
      \hfil}}%
\epsfxsize=0pt\epsfysize=0pt}%
%
%   We still need to define the tricky \epsfaux macro. This requires
%   a couple of magic constants for comparison purposes.
%
{\catcode`\%=12
\global\let\epsfpercent=%\global\def\epsfbblit{%BoundingBox}}%
%
%   So we're ready to check for `%BoundingBox:' and to grab the
%   values if they are found.
%
\long\def\epsfaux#1#2:#3\\{\ifx#1\epsfpercent
   \def\testit{#2}\ifx\testit\epsfbblit
      \epsfgrab #3 . . . \\%
      \epsffileokfalse
      \global\epsfbbfoundtrue
   \fi\else\ifx#1\par\else\epsffileokfalse\fi\fi}%
%
%   Here we grab the values and stuff them in the appropriate definitions.
%
\def\epsfgrab #1 #2 #3 #4 #5\\{%
   \global\def\epsfllx{#1}\ifx\epsfllx\empty
      \epsfgrab #2 #3 #4 #5 .\\\else
   \global\def\epsflly{#2}%
   \global\def\epsfurx{#3}\global\def\epsfury{#4}\fi}%
%
%   We default the epsfsize macro.
%
\def\epsfsize#1#2{\epsfxsize}%
%
%   Finally, another definition for compatibility with older macros.
%

% ================================================================
% execution: why not set
\epsfverbosetrue                        % reset at your pleasure
\abovefigskip=\baselineskip             % reset at your pleasure
\belowfigskip=\baselineskip             % reset at your pleasure
\global\let\figtitlefont\bf             % reset at your pleasure
\global\figtitleskip=.5\baselineskip    % reset at your pleasure

\font\tenmsb=msbm10   %%%% these first lines are to define \Bbb, for R etc.
\font\sevenmsb=msbm7
\font\fivemsb=msbm5
\newfam\msbfam
\textfont\msbfam=\tenmsb
\scriptfont\msbfam=\sevenmsb
\scriptscriptfont\msbfam=\fivemsb
\def\Bbb#1{\fam\msbfam\relax#1}
\let\nd\noindent %                                              NOINDENT
%                                          NEWLINE

\def\natural{{\rm I\kern-.18em N}}
 % or scaled\magstep2       TO INSERT LARGE
                            % FONT

\def\R{{\Bbb R}}

\def\chix{{\raise.5ex\hbox{$\chi$}}}
\def\chixa{{\chix\lower.2em\hbox{$_A$}}}

\def\real{{\rm I\kern-.2em R}}
\def\integer{{\rm Z\kern-.32em Z}}
\def\complex{\kern.1em{\raise.47ex\hbox{
            $\scriptscriptstyle |$}}\kern-.40em{\rm C}}
\def\vs#1 {\vskip#1truein}
\def\hs#1 {\hskip#1truein}

\def\Month{\ifcase\number\month \relax\or January \or February \or
  March \or April \or May \or June \or July \or August \or September
  \or October \or November \or December \else \relax\fi }
\def\date{\Month \the\day, \the\year}

  \hsize=6truein        \hoffset=.25truein %was \hoffset=1.2truein
  \vsize=8.8truein      %\voffset=1truein
  \pageno=1     \baselineskip=12pt
  \parskip=0 pt         \parindent=20pt
  \overfullrule=0pt     \lineskip=0pt   \lineskiplimit=0pt
  \hbadness=10000 \vbadness=10000 %     REPORT ONLY BEYOND THIS BADNESS
% Start of Text
%\nopagenumbers
%\lett
\pageno=0

\footline{\ifnum\pageno=0\hss\else\hss\tenrm\folio\hss\fi}
\hbox{}
\vskip 1truein\centerline{{\bf THE STRUCTURE OF THE HARD SPHERE SOLID}}
\vskip .2truein\centerline{by}
\vskip .2truein
\centerline{{Charles Radin
\footnote{*}{Research supported in part by NSF Grant DMS-0354994}
and Lorenzo Sadun
\footnote{**}{Research supported in part by NSF Grant DMS-0401655}}}
%\vskip .2truein\centerline{Department of Mathematics, University of Texas, Austin, TX}

\vs2
\centerline{{\bf Abstract}}
\vs.1 \nd
We show that near densest-packing the perturbations of the HCP
structure yield higher entropy than perturbations of any other densest
packing. The difference between the various structures shows up in the
correlations between motions of nearest neighbors. In the HCP
structure random motion of each sphere impinges slightly less on the
motion of its nearest neighbors than in the other structures.
\vs.8
\centerline{September 2004}
\vs.2
%\centerline{Subject Classification:\ \ 52A40, 52C26, 52C23}
\vfill\eject

\nd
{\bf I. Introduction.}

We are interested in the solid phase of the hard sphere gas model, a
phase which is generally agreed to exist based on computer experiments
refined over the past 50 years, as well as certain experiments with
monodisperse colloids. Although the existence of the solid phase is
uncontroversial, the internal structure of the solid is not well
understood, and is the object of this paper. See [1-2] for
reviews of the earliest computational work, showing the transition,
and [3-9] for more recent work trying to determine the internal
structure. See [10] for a relevant experiment with colloids.

The model consists of the classical statistical mechanics of point
particles for which the only interaction is a hard core: the
separation between particles must be at least 1. We use the canonical
ensemble, corresponding to fixed density $d$ and temperature $T$. (For
the remainder of the paper we use the more convenient terminology of
spheres rather than particles. In particular, by the ``density'' of a
configuration of spheres we will mean the fraction of space occupied
by the spheres.) In the usual way we can integrate out the velocity
variables and consider the ``reduced'' ensemble associated only with
the spatial variables. This ensemble is independent of temperature,
effectively leaving only the density variable $d$, and consists of the
uniform distribution on all configurations of the unit spheres at
density $d$. (For a finite system of $N$ spheres, constrained to lie
in a container $C\subset \R^3$ of volume $N/d$, a configuration can be
represented by the point in $C^N$ corresponding to the centers of the
spheres, and the uniform distribution is understood in the usual sense
of volume in $\R^{3N}$.) The entropy density of the finite-sphere
ensemble is then $S_{N,d}=(\log V_{N,d})/N$, where $V_{N,d}$ is
the subvolume of $C^N$ available to the (centers of the) spheres.

We will not be concerned with the solid/fluid transition, associated
with density around $0.54$, but with the nature of the solid near
maximum possible density, $d_c=\pi/\sqrt{18}\approx 0.74$.  The
configurations of density $d_c$ are known to be those obtained by
2-dimensional hexagonal layers, as follows. If we denote one
such layer by $\alpha$, then on either side of it we can choose either
of the two ways of ``filling the gaps'', either $\beta$ or
$\gamma$. The FCC lattice corresponds to the choice $\ldots, \alpha,
\beta,\gamma,\alpha, \beta,\gamma, \alpha, \beta,\gamma,\ldots$, the
HCP structure is obtained from the choice $\ldots,\alpha,
\beta,\alpha, \beta,\alpha, \beta, \ldots$, and there are infinitely
many other ``layered configurations'' of the same optimal
density. Since we will be concerned with an expansion of the ensemble
in the deviation $\Delta d =d_c-d$, there is a minor problem with
nonuniqueness of the configuration at the optimal density $d_c$. The
ensemble is, by construction, the distribution which maximizes
entropy. Our objective then is to show that, to lowest order in the
deviation $\Delta d$, perturbations of the HCP layering yield the
highest entropy compared with perturbations of other layerings. There
is computer evidence, and experimental evidence based on colloids,
that, however, it is the FCC layering which is optimal, by roughly the
same magnitude effect that we obtain, $0.1 \%$. We make many fewer assumptions than
these works, and will spell out our assumptions unambiguously.

The essential question is how much ``wiggle room'' is available to
each sphere.  To first approximation, this can be computed by freezing
the positions of all spheres but one, and computing the volume
available to the single unfrozen sphere.  FCC, HCP and all other close
packing configurations give exactly the same result to this order,
proportional to the volume of the Voronoi cell. The next approximation is to
consider the effect that the motion of one sphere has on the volume
available to its nearest neighbors.  To compute this effect, we freeze
the (equilibrium) positions of all but two nearest-neighbor spheres
and exactly compute the volume, in $\R^6$, of the allowed 2-sphere
configurations.  When the two nearest-neighbor spheres are in the same
layer, the results are the same for FCC and HCP or indeed any
layering. However, when the two spheres are in adjacent layers
there is slightly more available volume in the HCP case than in the
FCC or any other layering. We conclude that the
motion of each sphere in the HCP lattice impinges less on the motion
of its neighbors than the motion of each sphere in the FCC lattice,
and hence that small perturbations of the HCP lattice have more
entropy than small perturbations of the FCC lattice.

%A brief outline of our argument is as follows. Consider the volume of
%phase space obtained by freezing, in the positions of some layered
%configuration, the environment of pairs of neighboring spheres. (That
%is, starting from a perfect layered configuration at high density we
%free up isolated neighboring pairs of spheres.) We show that at high
%density this volume is optimized, among all possible layerings, by the
%HCP structure. 

\vs.2 \nd {\bf II. Calculations}

We choose Cartesian $(x,y,z)$ coordinates such that there are
hexagonal layers parallel to the $x,y$ plane. In particular we will
call that layer a $\beta$-plane which contains
sphere centers at the origin $O= (0,0,0)$ and the six sites:
$$\matrix{
a= (\sqrt{{1\over 4}}, \sqrt{{3\over 4}},0)
&b=(1,0,0)
&c=(\sqrt{{1\over 4}}, -\sqrt{{3\over 4}},0)\cr
d=(-\sqrt{{1\over 4}}, -\sqrt{{3\over 4}},0)
&e=(-1,0,0)
&f=(1\sqrt{{1\over 4}}, \sqrt{{3\over 4}},0)\cr} \eqno 1)$$

\nd See Figure 1. The centers for spheres in the layers above or below
this layer are possible at some of: 
$$\matrix{
A^{\pm}=(0,\sqrt{{1\over 3}}, \pm \sqrt{{2\over 3}})
&B^{\pm}=(\sqrt{{1\over 4}},\sqrt{{1\over 12}}, \pm\sqrt{{2\over 3}})\cr
C^{\pm}=(\sqrt{{1\over 4}},-\sqrt{{1\over 12}}, \pm\sqrt{{2\over 3}})
&D^{\pm}=(0,-\sqrt{{1\over 3}}, \pm \sqrt{{2\over 3}})\cr
E^{\pm}=(-\sqrt{{1\over 4}},-\sqrt{{1\over 12}}, \pm\sqrt{{2\over 3}})
&F^{\pm}=(-\sqrt{{1\over 4}},\sqrt{{1\over 12}}, \pm\sqrt{{2\over 3}})\cr} \eqno 2)$$

\nd See Figure 1. 

Consider the Voronoi cell of the sphere centered at $O$. Without
loss of generality, we assume there are spheres in the layer above
$O$, with $z$-coordinate of the centers equal to $2/3$, at sites
$A^+, C^+$ and $E^+$. We will call this an $\alpha$-plane. In the
layer below $O$, $z=-2/3$, there are spheres at either $A^-, C^-$
and $E^-$ (another $\alpha$-plane, for instance for HCP), or at
$B^-,D^-$ and $F^-$ (a $\gamma$-plane, for instance for FCC). In the
latter case the Voronoi cell is a rhombic dodecahedron, with the 14
vertices:
$$\matrix{
(\pm {1\over 2},-\sqrt{{1\over 12}},\sqrt{{1\over 6}})
&(\pm {1\over 2},\sqrt{{1\over 12}},\sqrt{{1\over 24}})\cr
(\pm {1\over 2},\sqrt{{1\over 12}},-\sqrt{{1\over 6}})
&(\pm {1\over 2},-\sqrt{{1\over 12}},-\sqrt{{1\over 24}})\cr} \eqno 3a)$$
$$\matrix{
(0,0,\sqrt{{3\over 8}})
&(0,\sqrt{{1\over 3}}, \sqrt{{1\over 6}})
&(0,\sqrt{{1\over 3}}, -\sqrt{{1\over 24}})\cr
(0,0,-\sqrt{{3\over 8}})
&(0,-\sqrt{{1\over 3}}, -\sqrt{{1\over 6}})
&(0,-\sqrt{{1\over 3}}, \sqrt{{1\over 24}})\cr} \eqno 3b)$$

\nd and in the former
case it is a trapezo-rhombic dodecahedron, with 14 vertices:
$$\matrix{
(\pm {1\over 2},\sqrt{{1\over 12}},\sqrt{{1\over 6}})
&(\pm {1\over 2},-\sqrt{{1\over 12}},\sqrt{{1\over 24}})\cr
(\pm {1\over 2},-\sqrt{{1\over 12}},-\sqrt{{1\over 24}})
&(\pm {1\over 2},\sqrt{{1\over 12}},-\sqrt{{1\over 6}})\cr} \eqno 4a)$$
$$\matrix{
(0,0,\sqrt{{3\over 8}})
&(0,-\sqrt{{1\over 3}}, \sqrt{{1\over 6}})
&(0,-\sqrt{{1\over 3}}, -\sqrt{{1\over 6}})\cr
(0,0,-\sqrt{{3\over 8}})
&(0,\sqrt{{1\over 3}}, -\sqrt{{1\over 24}})
&(0,\sqrt{{1\over 3}}, \sqrt{{1\over 24}})\cr} \eqno 4b)$$
\vs0
\epsfig .55\hsize; 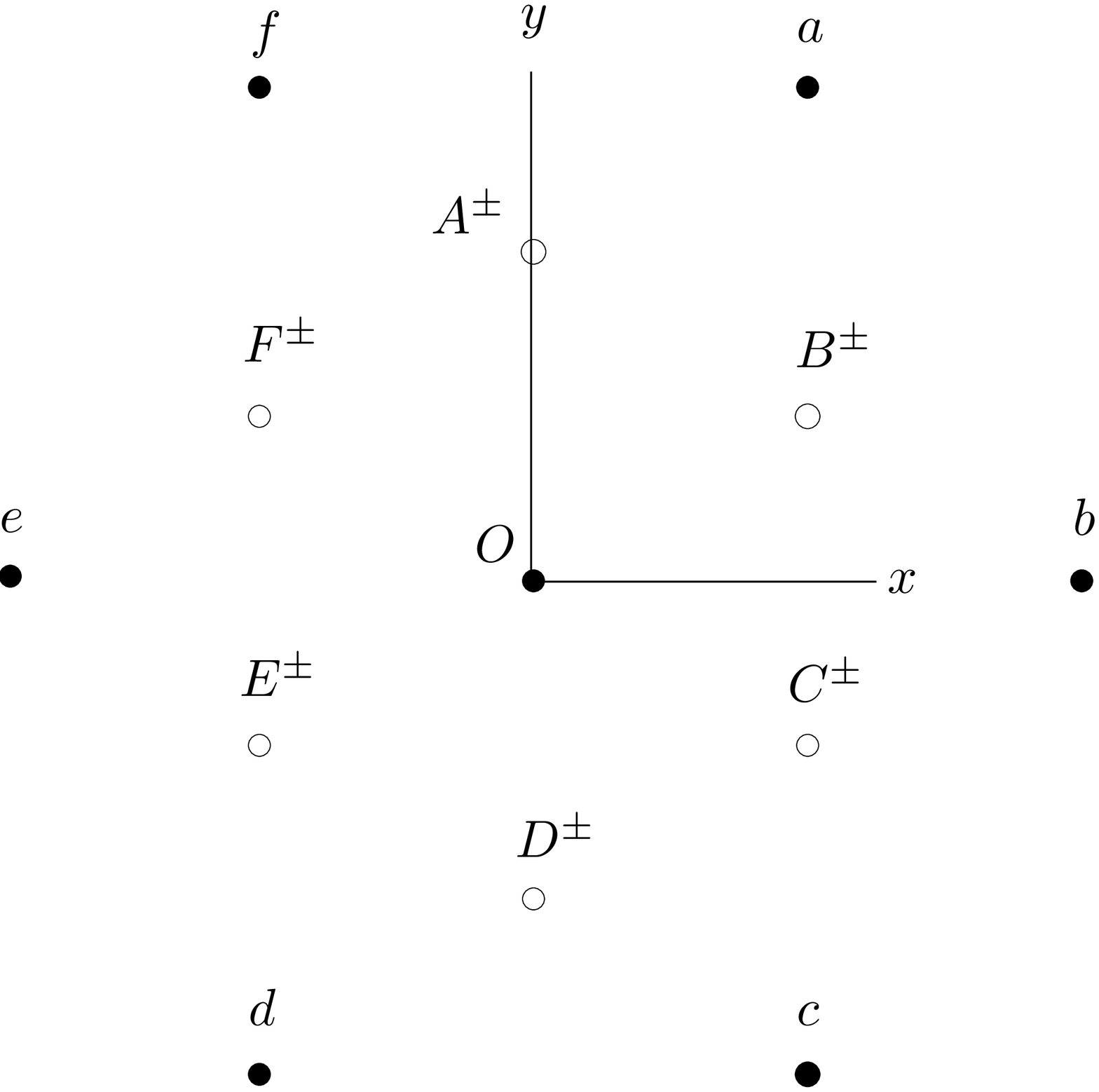;
\vs-.4
\nd Figure 1. 
$A^\pm, B^\pm,\ldots, F^\pm$ are centers of spheres above or below the $x-y$ plane;
$a,b,\ldots, f$, and $O$, are in the plane.
\vs.4

Neighboring spheres of diameter 1 centered at the sites 3) or 4)
actually touch. To deal with densities below $d_c$ it is convenient to
shrink the spheres rather than move the centers. Without loss of
generality we need only consider five cases of pairs of neighboring
spheres as follows. As noted above, the spheres in the $z=2/3$ plane
are an $\alpha$-plane, and those in the $z=0$ plane are a
$\beta$-plane. The five cases of pairs of spheres can then be chosen
with centers at: $O$ and $b$, with the $z=-2/3$ plane being
either $\alpha$ or $\gamma$; or centered at $O$ and $A^+$
with one of three possibilities: $\alpha$ for the $z=-2/3$ plane and
$\beta$ for the $z=4/3$ plane, or $\gamma$ for the $z=-2/3$ plane and
$\beta$ for the $z=4/3$ plane, or $\gamma$ for both the $z=-2/3$ plane
and the $z=4/3$ plane.  Again, our aim is to free up such a pair of
spheres, leaving their environment frozen in place, and compute the
volume in $\R^6$ available to the centers of the pair.

Imagine first that only the sphere at $O$ is freed up from its
lattice position, and consider the volume in $\R^3$ available for its
center. The boundary of this region consists of portions of 12
spherical surfaces -- think of the free sphere rolling on the surface
of its frozen neighbors. If the density is close to $d_c$ then the
region is very small, and to lowest order in $\Delta d$ we can
linearize these surfaces, obtaining a small copy -- volume of order
$(\Delta d)^3$ -- of the Voronoi cell of the (frozen) central sphere. Now
free up the other sphere also, the one centered at $A^+$. Then
each sphere is in part constrained by its 11 frozen neighbors, but
also by the other free sphere. Again to lowest order in $\Delta d$ we
can assume that the constraint on each free sphere due to the other
free sphere only depends on one degree of freedom, a coordinate along
the line separating their frozen centers. In other words, the region
available to one of the free spheres is, to lowest order in $\Delta
d$, the polyhedron obtained by moving inward or outward one of the
faces of the (small) Voronoi cell, simultaneously extending the faces
that touch the moving face. At maximum separation each sphere is then constrained
by an 11-sided polyhedron $\tilde P$. In Figure 2 we give an analogous
2-dimensional version of this process for a pair of circles freed up
from an hexagonal packing.

The (entropic) volume $V_S$ in $\R^6$ which we want to compute can then
be represented as:

$$V_S= \int_{B_1}^1 \Big[\int_{B_2}^{{1\over 2}-w}A_{\bar w}\,d\bar w\Big] A_w\,dw \eqno 5)$$

\nd where $A_w$ represents a cross-sectional area of one of these
maximal regions $\tilde P$, cut by a plane midway between the two spheres,
and $B_j$ refers to the end of $\tilde P_j$ which is opposite the other free sphere.

We have determined these cross-sectional areas, as follows. 
The polyhedron $\tilde P$ is associated with
the sphere near the origin, and there are four cases to consider:
whether the frozen configuration is FCC or HCP -- note that every other
layering would produce the same effect as one of these for this
computation -- and whether the second sphere is in the $z=0$ plane or 
the $z=2/3$ plane. The results are as follows.

For three of the cases, namely: FCC and the second sphere in the $z=0$
plane, FCC and the second sphere in the $z=2/3$ plane, and HCP and the
second sphere in the $z=0$ plane, we get the same results:

$${\pi^2\over 2(\Delta d)^2}
A_w=\cases{\sqrt{{1\over 32}}(6+8w),\ \ -{1\over 2}\le w\le 0\cr
\sqrt{{1\over 32}}(6-8w),\ \ 0\le w\le {1\over 2}\cr
\sqrt{{1\over 32}}(8-16w+8w^2),\ \ {1\over 2}\le w\le 1\cr}\eqno 6)$$

The fourth case is different, HCP and the second sphere in the $z=2/3$ plane:
$${\pi^2\over 2(\Delta d)^2}
A_w=\cases{\sqrt{{1\over 32}}(16+48w+36w^2),\ \ -{2\over 3}\le w\le -{1\over 2}\cr
\sqrt{{1\over 32}}({37\over 4}+21w+9w^2),\ \ -{1\over 2}\le w\le -{1\over 3}\cr
\sqrt{{1\over 32}}({21\over 4}-3w-27w^2),\ \ -{1\over 3}\le w\le -{1\over 6}\cr
\sqrt{{1\over 32}}({23\over 4}+3w-9w^2),\ \ -{1\over 6}\le w\le 0\cr
\sqrt{{1\over 32}}({23\over 4}-5w-9w^2),\ \ 0\le w\le {1\over 6}\cr
\sqrt{{1\over 32}}(6-8w),\ \ {1\over 6}\le w\le {1\over 2}\cr
\sqrt{{1\over 32}}(8-16w+8w^2),\ \ {1\over 2}\le w\le 1\cr}\eqno 7)$$
\vs.1
\nd We graph these two functions $A_w$ in Figure 3.

\hbox{}
\vs-.3
\hbox{}
\epsfig .5\hsize; 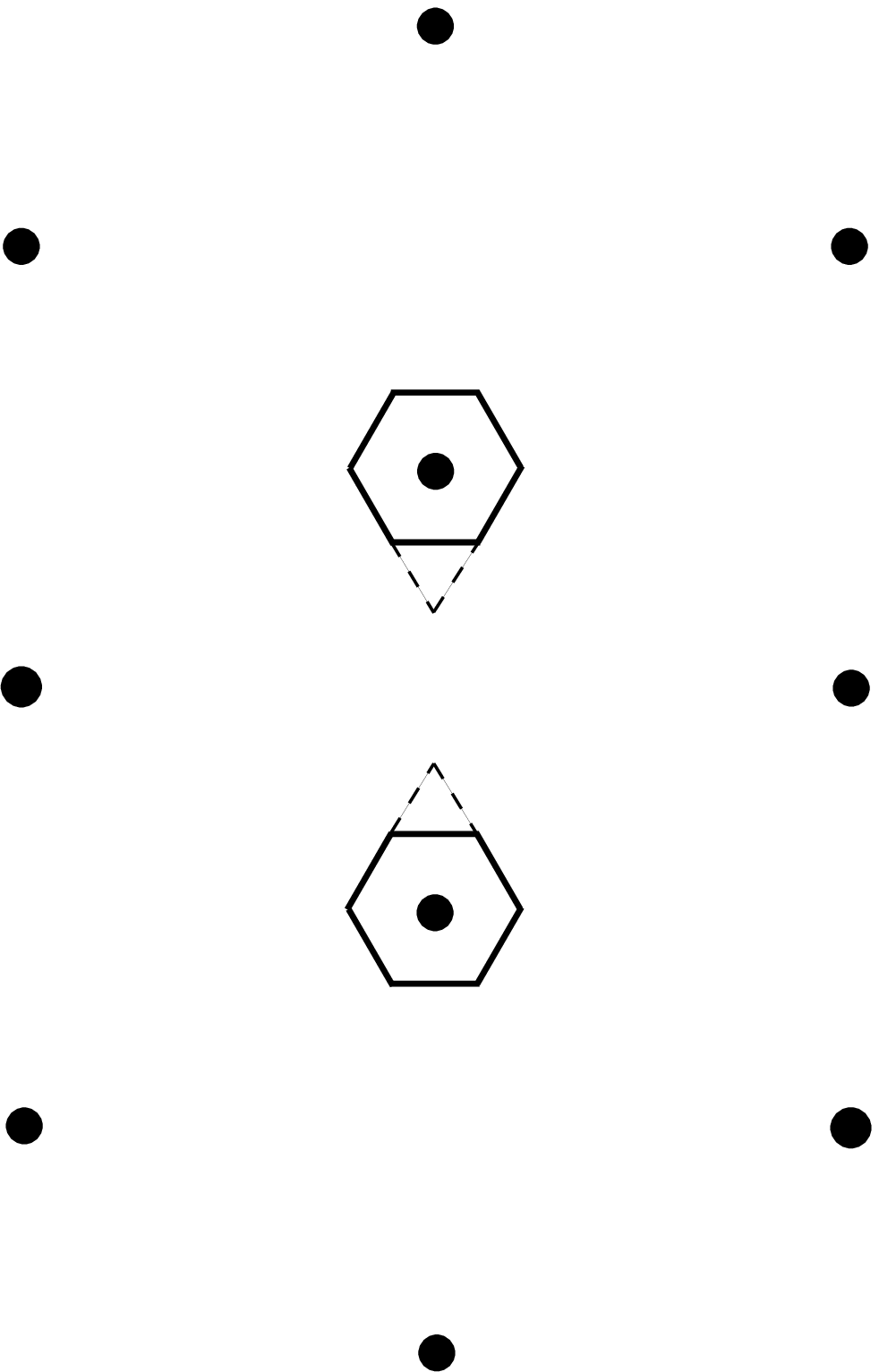;

\nd Figure 2. Small copies of the Voronoi cells of 2 disks, with
dashed lines showing how they extend when the centers of the disks separate.
\vfill \eject
\hbox{}
\vs-2
\epsfig .7\hsize; 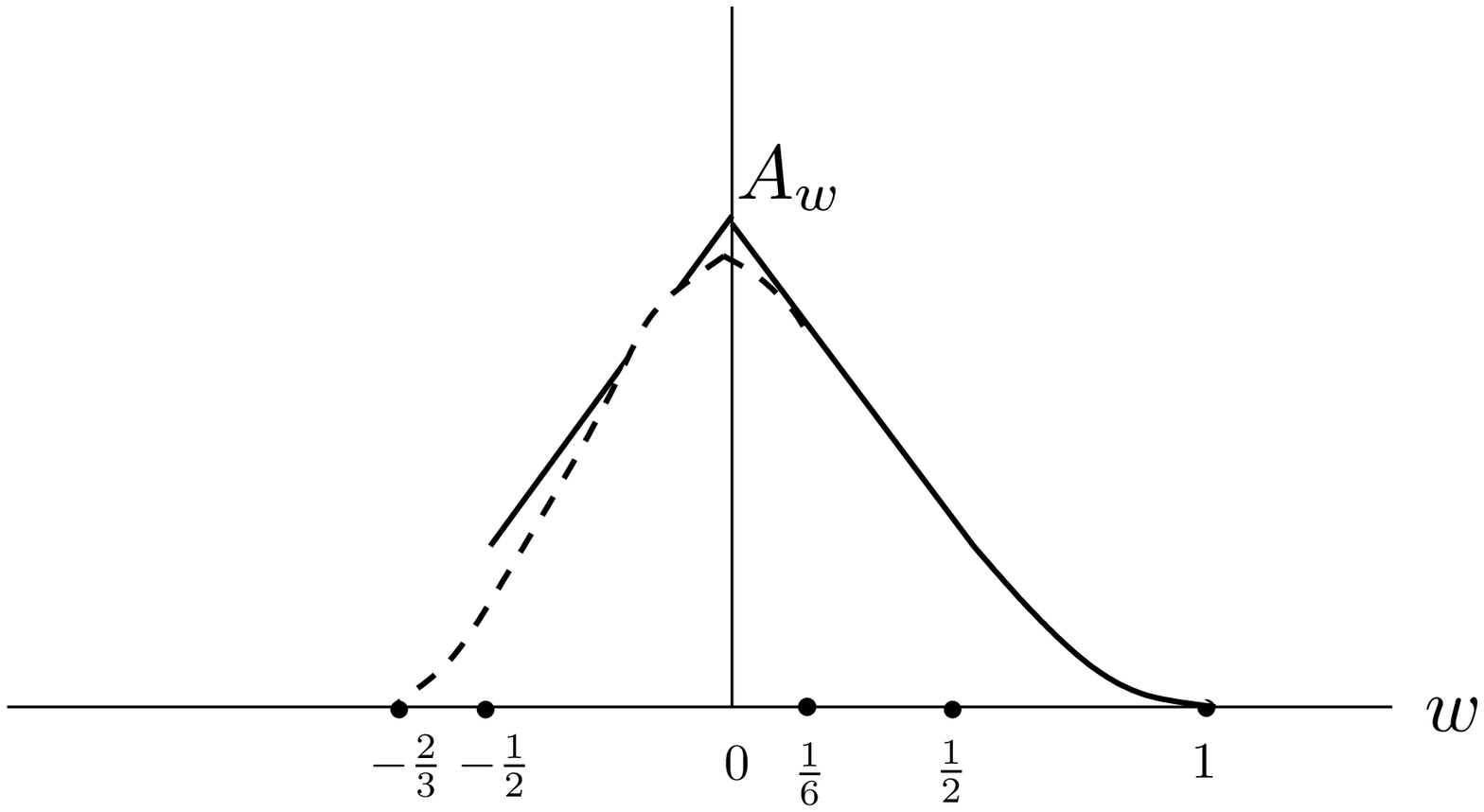;
\vs-2
\centerline{Figure 3. The 2 area functions of 6) and 7). They overlap
for $w\ge 1/6$.}
\vs.3

It only remains to compute $V_S$ from 5) for each of the five distinct
cases of pairs of neighboring spheres. We have done this and obtained
the following results. 

It is immediate from 6) that the two cases in which the second sphere
is also in the $z=0$ plane will have the same value, and that this
value will be the same if the $z=4/3$ and $z=-2/3$ planes are both
$\gamma$ -- for instance FCC; this value of $V_S$ is
$(467/960)[2(\Delta d)^3/\pi^2]\approx 0.48646[2(\Delta d)^3/\pi^2]$.

If the second sphere is in the $z=2/3$ plane, the $z=4/3$ plane is
$\beta$ and the $z=-2/3$ plane is $\alpha$ -- for instance HCP -- then
$V_S=(908179/1866240)[2(\Delta d)^3/\pi^2]\approx 0.48664[2(\Delta d)^3/\pi^2]$.

Finally, if the second sphere is in the $z=2/3$ plane, the $z=4/3$ plane is
$\beta$ and the $z=-2/3$ plane is $\gamma$, then
$V_S=(1814587/3732480)[2(\Delta d)^3/\pi^2]\approx 0.48616[2(\Delta d)^3/\pi^2]$.

These results prove our assertion on the optimality of the HCP
layering. They also allow us to quantify the entropy difference between
HCP and FCC.  Each off-layer ``bond'' in the HCP configuration has entropy 
$\log(908179/907848)$ 
greater than in the FCC configuration.  Half of this
difference is associated with each sphere.  However, each sphere has 6
nearest neighbors in different layers, so the HCP entropy per sphere is
$3 \log(908179/907848) \approx 0.0011$ greater than the entropy of
the FCC (and more for other layerings).

\vs.2 \nd {\bf III. Summary}

Our goal was to compare the entropies of certain families of
perturbations of the perfect densest packings of unit spheres. We
start with packings obtained from the densest packings, viewed as
consisting of two dimensional hexagonal layers, by homogeneously
lowering the density -- for instance by uniformly shrinking the size
of the spheres. From these various starting points -- namely the
various layerings, including FCC and HCP, which are lower density
versions of the densest packings -- we make two assumptions. First we
look only for terms of lowest order in the deviation of density from
densest packing. And second, we only consider those perturbations
obtained by loosening isolated pairs of neighboring spheres from their
lattice positions. Clearly the latter is our only nontrivial
assumption. Our result is that perturbations of the HCP structure have
the largest entropy, in contradiction with [9] in which it is claimed
that the contributions from nearest neighbor spheres alone yields a
preference for FCC. 

\vs.2 \nd {\bf IV. References}
\vs.2
\item{[1]} J.A. Barker, {\it Lattice Theories of the Liquid State},
Macmillan, New York, 1963.

\item{[2]} H.N.V. Temperley, J.S. Rowlinson and G.S. Rushbrooke,
{\it Physics of Simple Liquids}, Wiley, New York, 1968.

\item{[3]} D. Frenkel and B. Smit, {\it Understanding Molecular
Simulation: From Algorithms to Applications}, Academic, Boston, 1996.

\item{[4]} A.D. Bruce, N.B. Wilding and G.J. Ackland, Free Energy of
Crystalline Solids: A Lattice-Switch Monte Carlo Method, {\it
Phys. Rev. Lett.} {\bf 79}, 3002, 1997.

\item{[5]} L.V. Woodcock, {\it Nature} (London) {\bf 385}, 141-143, 1997.

\item{[6]} L.V. Woodcock, {\it Nature} (London) {\bf 388}, 236-237,
1997.

\item{[7]} R.J. Speedy, Pressure and entropy of hard-sphere crystals,
{\it J. Phys: Condens. Matter} {\bf 10}, 4387-4391, 1998.

\item{[8]} P.G. Bolhuis, D. Frenkel, S.-C. Mau and D.A. Huse,
{\it Nature} {\bf 388}, 235-236, 1997.

\item{[9]} S.-C. Mau and D.A. Huse, Stacking entropy of hard-sphere
crystals, {\it Phys. Rev. E} {\bf 59}, 4396-4401, 1999.

\item{[10]} P.N. Pusey et al, Structure of crystals of hard colloidal
spheres, {\it Phys. Rev. Lett.} {\bf 63}, 2753-2756, 1989.

\vs.2 \nd
{{\bf Acknowledgments}} 
\vs.1
It is a pleasure to thank the Aspen Center for Physics for support at
the Workshop on Geometry and Materials Physics in June 2004, and to
thank Randy Kamien for pointing us to the paper by Mau and Huse.

\vs.5 \nd
\line{Charles Radin, Mathematics Department, University of Texas, Austin, TX\ \ 78712}

\nd {\it Email address}: {\tt radin@math.utexas.edu}
\vs.05 \nd
\line{Lorenzo Sadun, Mathematics Department, University of Texas, Austin, TX\ \ 78712}

\nd {\it Email address: {\tt sadun@math.utexas.edu}}
\vfill 

\end